\documentclass[12pt]{article}

\usepackage{amsmath,amssymb,cite}

\def\tablerule{\noalign{\hrule}}

\def\be{\begin{equation}}
\def\eqn{\begin{equation}\label}
\newcommand{\ee}{\end{equation}}

\def\bea{\begin{eqnarray}}
\def\eqnn{\bea\label}
\def\eea{\end{eqnarray}}
\def\nn{\nonumber}

\def\md{\medskip}

\newcommand{\eqna}[1]{\begin{subequations} \label{#1}
\begin{eqnarray}}
\def\eena{\end{eqnarray}
\end{subequations}}

 \def\nt{\noindent}
\def\nl{\hfill\break}

\def\nlii{\\[3pt]\indent}

\def\half{{\textstyle{1\over2}}}
\def\ha{{\textstyle{1\over2}}}

\def\halfn{{\textstyle{N\over 2}}}
\def\quartn{{\textstyle{N\over 4}}}

\def\novert{{\textstyle{2\over N}}}

\def\quartern{{\textstyle{N\over4}}}

\def\eighth{{\textstyle{1\over8}}}

\def\quarter{{\textstyle{1\over4}}}

\def\third{{\textstyle{1\over3}}}

\def\rg{\rangle} 

\font\fat=cmsy10 scaled\magstep5

\def\Bbullet{\raise-3pt\hbox{\fat\char"0F}}

\def\tV{\tilde V} 

 \def\hL{\hat L}

\def\dia{~~$\diamondsuit$}

\def\bu{$\bullet$}
\def\buu{\bullet\ }

\def\cg{{\cal G}} \def\ch{{\cal H}} 
  
\def\cm{{\cal M}}

\def\white#1{\mathop{\bigcirc}\limits_{#1}}
\def\gray#1{\mathop{\bigotimes}\limits_{#1}}
\def\riga{-\kern-4pt - \kern-4pt -}

\def\bbz{Z\!\!\!Z}
\def\bbc{C\kern-6pt I}
\def\bac{{C\kern-5.5pt I}}

\def\bbn{I\!\!N} \def\k{\kappa} \def\o{{\bar 0}} \def\I{{\bar 1}}

\def\a{\alpha} \def\b{\beta} \def\g{\gamma} \def\d{\delta}

\def\D{\Delta} \def\L{\Lambda}

\def\Ra{\Longrightarrow}

\def\({\left(}
\def\){\right)}

\def\a{\alpha}
\def\b{\beta}
\def\d{\delta}
\def\g{\gamma}

\def\D{\Delta}
\def\L{\Lambda}

\begin{document}

\hfill SISSA/86/2010/EP

\hfill ESI 2289 (2010)

\hfill hep-th/1012.3685

\begin{center}

{\LARGE {\bf Group-Theoretical\\[2pt]
 Classification of BPS and\\[2pt]
  Possibly Protected States in D=4\\[3pt]
  Conformal Supersymmetry}
}

 \vspace{10mm}

{\bf \large V.K. Dobrev}

\vskip 5mm

\emph{Scuola Internazionale Superiore di Studi Avanzati,\\
via Bonomea 265, 34136 Trieste, Italy,\\
\vskip 4mm   \vskip 4mm Erwin Schr\"odinger International Institute
for Mathematical Physics,
Boltzmanngasse 9, A-1090 Vienna, Austria\\
\vskip 4mm and\\ \vskip 4mm
  Institute for Nuclear
Research and Nuclear Energy,\footnote{Permanent address.}\\
 Bulgarian Academy of Sciences,\\ 72
Tsarigradsko Chaussee, 1784 Sofia, Bulgaria}

\end{center}

\vskip 4mm

\begin{abstract}
{\small  We give explicitly the  reduction of supersymmetries of the positive
energy unitary irreducible representations of the N-extended D=4
conformal superalgebras  su(2,2/N). Further we give the classification of BPS and
   possibly protected states.}
\end{abstract}

\newpage

\section{Introduction}
\setcounter{equation}{0}

Recently, superconformal field theories in various dimensions are
attracting more interest, especially in view of their applications in string theory.
From these very important is  the AdS/CFT correspondence,
namely, the remarkable proposal of Maldacena \cite{Malda}, according to which the large
$N$ limit of a conformally invariant theory in $d$ dimensions is
governed by supergravity (and string theory) on $d+1$-dimensional
$AdS$ space (often called $AdS_{d+1}$) times a compact manifold.
Actually the possible relation of field theory on $AdS_{d+1}$ to
field theory on $\cm_d$ has been a subject of long interest, cf.,
e.g., \cite{FlFr,NiSe,GNW}, and also \cite{FeFr} for discussions
motivated by recent developments. The
proposal of \cite{Malda} was elaborated in \cite{GKP} and
\cite{Wi} where was proposed a precise correspondence between
conformal field theory observables and those of supergravity.
More recently, there were developments of  integrability in the  context of the AdS/CFT correspondence,
in which superconformal field theories, especially in 4D, were also playing important role. For this we refer to the reviews
\cite{Beisert,Torrielli}, and for earlier relevant papers also in the general context of AdS/CFT and superconformal symmetry,
we refer to \cite{Andrianopoli,Gava,Distler,Frampton,Gunaydin,Dobrev,Leonhardt,Ivanov,Yoneya,Gomez,Grigoriev,Giribet,Bhattacharya,Mas,Cornalba,Gubser}
and references therein.

In all cases, it was known for a long time that  the classification of the UIRs of the conformal
 superalgebras  is  of great importance.
For some time  such classification was known only for the ~$D=4$~
superconformal algebras ~$su(2,2/1)$ \cite{FF} and ~$su(2,2/N)$~ for
arbitrary $N$ \cite{DPu}, (see also \cite{DPm,DPf}).
 Then, more progress was made with the classification
for ~$D=3$ (for even $N$), $D=5$, and $D=6$ (for $N=1,2$)~
in \cite{Min} (some results being conjectural), then for the $D=6$
case (for arbitrary $N$) was finalized in \cite{Dosix}. Finally, the
cases $D=9,10,11$ were treated by finding the UIRs of $osp(1/2n)$,
\cite{DoZh}.

After the list of UIRs is found the next problem to address is to find their
characters since these give the spectrum which is important for the
applications. This problem is solved in principle, though not all formulae are explicit,
for the UIRs of $D=4$ conformal superalgebras $su(2,2/N)$ in \cite{Doch}.\footnote{For another,
more practical though not so rigorous, approach, see \cite{BDHO}.}
From the mathematical point of view this question is clear only
for representations with conformal dimension above the unitarity
threshold viewed as irreps of the corresponding complex superalgebra
$sl(4/N)$ \cite{BL,JHKT,Jeu,Serga,VZ,Bru,SuZh}.  But for $su(2,2/N)$ even the UIRs above the unitarity
threshold are truncated for small values of spin and isospin. Furthermore,
 in the applications the most important role is played by
the representations with ``quantized" conformal dimensions at the
unitarity threshold and at discrete points below.  In the quantum
field or string theory framework some of these correspond to
operators with ``protected" scaling dimension and therefore imply
``non-renormalization theorems" at the quantum level, cf., e.g.,
\cite{HeHo,FSa}. Especially important in this context are the
so-called BPS states, cf.,
\cite{AFSZ,FSb,FSa,EdSo,Ryz,DHRy,ArSo,DHHR,Dolan}.

These investigations require deeper knowledge of the structure of the UIRs,
in particular,   more explicit results on the decompositions of long superfields
as they descend to the unitarity threshold .    Fortunately,
most of the needed  information is contained in \cite{DPu,DPm,DPf,DPp,Doch},
see also \cite{FGPW,FS,AES,BKRSa,HeHoa,DoOs}.

The paper is organized as follows. In Section 2 we give the preliminaries.
In Section 3 we give explicitly the  reduction of supersymmetries.
In Section 4 we give the classification of BPS and
   possibly protected states.

\newpage

\section{Preliminaries}
\setcounter{equation}{0}

\subsection{Representations of D=4 conformal supersymmetry}

 The conformal superalgebras in $D=4$ are ~$\cg ~=~ su(2,2/N)$.
The even subalgebra of ~$\cg$~ is the algebra ~$\cg_0 ~=~ su(2,2)
\oplus u(1) \oplus su(N)$. We label their physically relevant
representations of ~$\cg$~ by the signature: \eqn{sgn}\chi ~=~
[\,d\,;\, j_1\,,\,j_2\,;\,z\,;\,r_1\,,\ldots,r_{N-1}\,]\end{equation}  where
~$d$~ is the conformal weight, ~$j_1,j_2$~ are non-negative
(half-)integers which are Dynkin labels of the finite-dimensional
irreps of the $D=4$ Lorentz subalgebra ~$so(3,1)$~ of dimension
~$(2j_1+1)(2j_2+1)$, ~$z$~ represents the ~$u(1)$~ subalgebra which
is central for ~$\cg_0$~ (and  is central for $\cg$
itself when $N=4$), and ~$r_1,\ldots,r_{N-1}$~ are non-negative integers which
are Dynkin labels of the finite-dimensional irreps of the internal
(or $R$) symmetry algebra ~$su(N)$.

We  recall the root system of the complexification ~$\cg^\bac$~ of
$\cg$  (as used in \cite{DPf}). The positive root system ~$\D^+$~ is
comprised of ~$\a_{ij}\,$, ~$1\leq i <j \leq 4+N$. The even positive
root system ~$\D^+_\o$~ is comprised of ~$\a_{ij}\,$, with\ $i,j\leq
4$~ and ~$i,j\geq 5$; ~the odd positive root system ~$\D^+_\I$~ is
comprised of ~$\a_{ij}\,$, with ~$i\leq 4, j \geq 5$. The generators corresponding
to the latter (odd) roots will be denoted as ~$X^+_{i,4+k}\,$, where ~$i=1,2,3,4$,
~$k=1,\ldots,N$.
The simple roots are chosen as in (2.4) of \cite{DPf}: \eqn{smplr}
\g_1 = \a_{12}\ , ~\g_2 = \a_{34}\ , ~\g_3 = \a_{25}\ , ~\g_4 =
\a_{4,4+N}\ , ~~\g_k = \a_{k,k+1}\ , ~~ 5\leq k\leq
3+N.\end{equation} Thus, the Dynkin diagram is: \eqn{dynk}
\vbox{\offinterlineskip\baselineskip=10pt \halign{\strut# \hfil &
#\hfil \cr &\cr & $\white{{1}} \riga \gray{{3}} \riga \white{{5}}
\riga \cdots \riga \white{{3+N}} \riga \gray{{4}} \riga \white{{2}}$
\cr }}\end{equation} This is a non-distinguished simple root system
with two odd simple roots \cite{Kacs}.

Sometimes we shall use another way of writing the signature related to the above
enumeration of simple roots, cf. \cite{DPf} and (1.16) of \cite{Doch}:
\eqn{signz} \chi ~=~ (2j_1\, ;\, (\L,\g_3)\,;\, r_1,\ldots,r_{N-1}\,;\,(\L,\g_4)\,;\,2j_2)\ ,\ee
(where ~$(\L,\g_3)$, $(\L,\g_4)$ are definite linear combinations of all quantum numbers),
or even
giving only the Lorentz and $SU(N)$ signatures:
\eqn{signzz} \chi_N ~=~ \{\,2j_1\, ;\, r_1,\ldots,r_{N-1}\,;\,2j_2\,\} \ .\ee

\nt{\bf Remark:}~~ We recall that the group-theoretical approach to
$D=4$ conformal supersymmetry developed in \cite{DPm,DPf,DPu}
involves two related constructions - on function spaces and as Verma
modules. The first realization employs the explicit construction of
induced representations of ~$\cg$~ (and of the corresponding
supergroup ~$G ~=~ SU(2,2/N)$) in spaces of functions (superfields)
over superspace which are called elementary representations (ER).
The UIRs of $\cg$ are realized as irreducible components of ERs, and
then they coincide with the usually used superfields in indexless
notation. The Verma module realization is also very useful as it
provides simpler and more intuitive picture for the relation between
reducible ERs, for the construction of the irreps, in particular, of
the UIRs. For the latter the main tool is an adaptation of the
Shapovalov form \cite{Sha} to the Verma modules \cite{DPu,DPp}. Here
we shall need only the second - Verma module - construction.\dia

We use lowest weight Verma modules $V^\L$ over $\cg^\bac$,
where the lowest weight $\L$ is characterized by its values on the Cartan
subalgebra ~$\ch$~ and is in 1-to-1 correspondence with the
signature $\chi$.
If a Verma module ~$V^\L$~ is irreducible then it gives the lowest
weight irrep ~$L_\L$~ with the same weight. If a Verma module
~$V^\L$~ is reducible then it contains a maximal invariant submodule
~$I^\L$~ and the lowest weight irrep ~$L_\L$~ with the same weight
is given by factorization: ~$L_\L ~=~ V^\L\,/\,I^\L$ \cite{Kacl}.

There are submodules which are generated by the singular vectors
related to the even simple roots
~$\g_1,\g_2,\g_5,\ldots,\g_{N+3}$~\cite{DPf}.
These generate an even invariant submodule $I^\L_c$
present in all Verma modules that we consider and which must
be factored out. Thus, instead of ~$V^\L$~ we shall
consider the factor-modules: \eqn{fcc} \tV^\L ~=~ V^\L\, /\, I^\L_c
\end{equation}

The Verma module reducibility conditions for the ~$4N$~
odd positive roots of $\cg^\bac$ were derived in
\cite{DPm,DPf} adapting the results of Kac \cite{Kacl}:
\eqna{redu} && d ~=~ d^1_{Nk} - z \d_{N4}
\\ &&  d^1_{Nk} ~\equiv~ 4-2k +2j_2 +z+2m_k -2m/N \cr
&&\cr && d ~=~ d^2_{Nk} - z \d_{N4} \\ &&  d^2_{Nk} ~\equiv~ 2-2k
-2j_2 +z+2m_k -2m/N
\cr &&\cr && d ~=~ d^3_{Nk} + z \d_{N4}\\
&& d^3_{Nk} ~\equiv~ 2+2k-2N +2j_1 -z-2m_k +2m/N \cr &&\cr && d ~=~
d^4_{Nk} + z \d_{N4}\\ && d^4_{Nk} ~\equiv~ 2k-2N -2j_1 -z-2m_k
+2m/N
\nn\eena \nt where in all four cases of (\ref{redu})
~$k=1,\ldots,N$, ~$m_N\equiv 0$, and \eqn{mkm} m_k \equiv
\sum_{i=k}^{N-1} r_i \ , \quad m \equiv \sum_{k=1}^{N-1} m_k =
\sum_{k=1}^{N-1} k r_k\end{equation}
Note that we shall use also the quantity $m^*$ which is conjugate to $m$~:
\eqnn{conjm}  m^* ~&\equiv&~ \sum_{k=1}^{N-1} k r_{N-k} ~=~ \sum_{k=1}^{N-1} (N-k) r_{k}\ ,\\
&& m+m^* ~=~ Nm_1 \ .\eea

We need  the result of \cite{DPu} (cf. part (i) of the
Theorem there) that the following is the complete list of lowest
weight (positive energy) UIRs of $su(2,2/N)$~: \eqna{unitt} && d
~\geq~ d_{\rm max} ~=~ \max (d^1_{N1}, d^3_{NN})\ ,\\ && d ~=~
d^4_{NN} \geq d^1_{N1}\ , ~~ j_1=0\ ,\\ && d ~=~ d^2_{N1} \geq
d^3_{NN}\ , ~~ j_2=0\ ,\\ && d ~=~ d^2_{N1} = d^4_{NN}\ , ~~ j_1 =
j_2=0\ ,\eena \nt where ~$d_{\rm max}$~ is the threshold of the
continuous unitary spectrum.
Note that in case (d) we have $d=m_1$, $z=2m/N -m_1\,$,
and that it is trivial for $N=1$.

Next we note that if ~$d ~>~ d_{\rm max}$~ the factorized Verma
modules are irreducible and coincide with the UIRs ~$L_\L\,$. These
UIRs are called ~${\bf long}$~ in the modern literature, cf., e.g.,
\cite{FGPW,FS,FSa,AES,BKRSa,EdSo,HeHoa}. Analogously, we shall use
for the cases when ~$d= d_{\rm max}\,$, i.e., (\ref{unitt}a), the
terminology of ~{\bf semi-short}~ UIRs, introduced in
\cite{FGPW,FSa}, while the cases (\ref{unitt}{b,c,d}) are also
called ~{\bf short}~ UIRs, cf., e.g.,
\cite{FS,FSa,AES,BKRSa,EdSo,HeHoa,DoOs}.

Next consider in more detail the UIRs at the four distinguished
reducibility points determining the UIRs list above: ~$d^1_{N1}\,$,
~$d^2_{N1}\,$, ~$d^3_{NN}\,$, ~$d^4_{NN}\,$.
The above reducibilities occur for the following odd roots, resp.:
\eqn{disr} \a_{3,4+N}~=~ \g_2 + \g_4  \ , \quad \a_{4,4+N} ~=~ \g_4
\ , \quad \a_{15} ~=~ \g_1 + \g_3 \ , \quad \a_{25}~=~ \g_3  \ . \end{equation}
We note a partial ordering of
these four points: \eqn{parto} d^1_{N1} ~>~ d^2_{N1} \ , \qquad
d^3_{NN} ~>~ d^4_{NN} \ .\end{equation}
 Due to this ordering ~{\it at most two}~ of these four points may coincide.

First we consider the situations in which ~{\it no two}~ of the
distinguished four points coincide. There are four such situations:
\eqna{dist} &{\bf a:} & ~~~
 d = d_{\rm max} = d^1_{N1} =
d^a \equiv 2 +2j_2 +z+2m_1 -2m/N > d^3_{NN}\qquad\\
& {\bf b:} &  ~~~d
~=~ d^2_{N1} = d^b \equiv z -2j_2 +2m_1 -2m/N  > d^3_{NN}\ , ~~ j_2=0\  \\
& {\bf c:} &~~~
 d ~=~ d_{\rm max} ~=~
d^3_{NN} ~=~ d^c ~\equiv~ 2+2j_1 -z +2m/N  > d^1_{N1}\  \\
&{\bf d:} &~~~
 d ~=~ d^4_{NN} = d^d \equiv 2m/N-2j_1 -z  > d^1_{N1}\ , ~~ j_1=0\
\eena
\noindent where for future use we have introduced notations ~$d^a,d^b,d^c,d^d$,
the definitions including also the corresponding inequality.

We shall call these cases ~{\bf single-reducibility-condition
(SRC)}~ Verma modules or UIRs, depending on the context.
In addition, as already stated, we use for the cases when ~$d= d_{\rm
max}\,$, i.e., (\ref{dist}a,c),
the terminology of semi-short UIRs,
while the cases (\ref{dist}b,d),
are also called short UIRs.

The factorized Verma modules ~$\tV^\L$~ with the unitary signatures
from (\ref{dist}) have only one invariant odd submodule which has
to be factorized in order to obtain the UIRs. These odd embeddings
and factorizations are given  as follows:
\eqn{embs}\tV^\L ~\rightarrow~ \tV^{\L+\b}\ , \qquad
L_\L ~=~ \tV^\L/I^\b \ , \end{equation}
where we use the convention \cite{DPm} that arrows point to the
oddly embedded module, and we give only the cases for ~$\b$~ that we shall
use later:
\eqna{paorz} \b ~&& =~ \a_{3,4+N}\ , \quad {\rm for}~ (\ref{dist}a),
\quad j_2> 0 ,\\
&&  =~
\a_{3,4+N}+\a_{4,4+N}\ , \quad {\rm for}~ (\ref{dist}a), \quad
j_2=0, \\
&&  =~ \a_{15}\ ,\quad {\rm for}~ (\ref{dist}c), \quad j_1>
0,\\
 &&  =~ \a_{15}+\a_{25}\ , \quad {\rm for}~
(\ref{dist}c),\quad j_1=0 \quad
\eena

We consider  now the four situations in which ~{\it two}~
distinguished points coincide: \eqna{disd}
&{\bf ac:} &\quad
d ~=~ d_{\rm max}
~=~ d^{ac} ~\equiv~ 2 + j_1 + j_2 + m_1 =
d^1_{N1} = d^3_{NN}\\
&{\bf ad:} &\quad  d ~=~ d^{ad} ~\equiv~  1 + j_2 + m_1 ~=~ d^1_{N1}
=
d^4_{NN}  \ , ~~ j_1=0 \qquad\\
&{\bf bc:}
&\quad  d ~=~ d^{bc} ~\equiv ~ 1 + j_1 + m_1 ~=~ d^2_{N1} = d^3_{NN}  \ , ~~ j_2=0 \\
&{\bf bd:} &\quad  d ~=~ d^{bd} ~\equiv ~ m_1 ~=~ d^2_{N1} =
d^4_{NN}  \ , ~~ j_1=j_2=0 \eena \nt We shall call these ~{\bf
double-reducibility-condition (DRC)} Verma modules or UIRs. The
cases in (\ref{disd}a) are semi-short UIR, while the other cases are
short.

\nt The odd embedding diagrams and factorizations
for the DRC modules are \cite{DPm}:
\eqn{embd} \vbox{\qquad $\begin{matrix}
\tV^{\L+\b'}&\rightarrow & \tV^{\L+\b+\b'}  \cr
&&\cr
\uparrow &&\uparrow  \cr
&&\cr
\tV^\L &\rightarrow & \tV^{\L+\b} \cr
\end{matrix}$ } \ee
$ L_\L ~=~ \tV^\L
/I^{\b,\b'} \ , \quad I^{\b,\b'} ~=~ I^\b \cup I^{\b'}$\\
and we give only the cases for ~$\b,\b'$~ to be used later:
\eqna{paor}  (\b,\b')   ~&=&~ (\a_{15},\a_{3,4+N}), \quad {\rm for}~
(\ref{disd}a), \quad j_1j_2> 0\\
~&=&~ (\a_{15},\a_{3,4+N}+\a_{3,4+N}), \quad {\rm for}~
(\ref{disd}b),\quad j_1> 0,\ j_2=0\\
 ~&=&~ (\a_{15}+\a_{25},\a_{3,4+N}),\quad {\rm for}~ (\ref{disd}c), \quad
j_1=0,\ j_2> 0\\
 ~&=&~ (\a_{15}+\a_{25},\a_{3,4+N}+\a_{3,4+N}),
~~ {\rm for}~ (\ref{disd}d),~~  j_1=j_2=0\qquad\quad
 \eena

\subsection{Decompositions of long superfields}
\label{deco}

First we present  the results on decompositions of long
irreps as they descend to the unitarity threshold \cite{Doch}.

In the SRC cases
we have established that for ~$d=d_{\rm max}$~ there hold the
 two-term decompositions:
\eqn{decog}
\(\hL_{\rm long}\)_{\vert_{d=d_{\rm max}}}
 ~=~ \hL_\L ~\oplus \hL_{\L+\b}
\ , \qquad r_1 + r_{N-1} >0\ ,\end{equation}
 where $\L$ is a semi-short SRC
designated as type {\bf a} (then $r_1>0$)
or {\bf c} (then $r_{N-1}>0$)
and there are four possibilities for
$\b$ depending on the values of $j_1,j_2$
as given in (\ref{paorz}).
In  cases (\ref{paorz}a,c)
also the second UIR on the RHS of (\ref{decog})
is semi-short, while in cases (\ref{paorz}b,d)
the second UIR on the RHS of (\ref{decog})
is short of type {\bf b}, {\bf d}, resp.

In the DRC cases    we have
established that for $N>1$ and ~$d=d_{\rm max}=d^{ac} $~ hold the
four-term decompositions:
\eqn{decggg} \(\hL_{\rm long}\)_{\vert_{d=d^{ac}}} ~=~ \hL_{\L}
~\oplus~ \hL_{\L +\b} ~\oplus~ \hL_{\L + \b'} ~\oplus~ \hL_{\L +\b+
\b'}\ , \qquad r_1r_{N-1} >0\ ,\end{equation}
 where $\L$ is the semi-short DRC
designated as type {\bf ac} and there are four possibilities for
$\b,\b'$ depending on the values of $j_1,j_2$
as given in (\ref{paor}a,b,c,d).
Note that in  case (\ref{paor}a)
all UIRs in the RHS of (\ref{decggg})
are semi-short.
In the case (\ref{paor}b)
the first two UIRs in the RHS of (\ref{decggg})  are
semi-short, the last two UIRs
are short of type {\bf bc}.
In the case (\ref{paor}c)
the first two UIRs in the RHS of (\ref{decggg})  are
semi-short, the last two UIRs
are short of type {\bf ad}.
In the case (\ref{paor}d)
the first UIR  in the RHS of (\ref{decggg})
is semi-short, the other three UIRs are
short of types {\bf bc}, {\bf ad}, {\bf bd}, resp.

Next we note  that for $N=1$ all SRC cases
enter some decomposition, while no DRC cases enter any
decomposition. For $N>1$ the situation is more
diverse and so we give the list of UIRs that do {\bf not} enter
decompositions together with the restrictions on the $R$-symmetry
quantum numbers:

\md

\nt\bu ~~~{\bf SRC cases:}

\nt\bu{\bf a} ~~~
$d ~=~   d^a   \ , \qquad r_1=0$\ .

\nt{\bu{\bf b}}~~~
$d ~=~  d^b   \ , \qquad r_1\leq 2$\ .

\nt\bu{\bf c} ~~~
$d ~=~ d^c   \ , \qquad  r_{N-1}=0$\ .

\nt{\bu{\bf d}}~~~
$d ~=~ d^d   \ , \qquad r_{N-1}\leq 2$\ .

\md

\nt\bu ~~~{\bf DRC cases:}\nl
all non-trivial cases for $N=1$, while for $N>1$ the list is:

\nt\bu{\bf ac} ~~~
$d ~=~   d^{ac} \ , \qquad r_1r_{N-1}=0$\ .

\nt{\bu{\bf ad}}~~~
$d ~=~ d^{ad} \ , \qquad
    r_{N-1}\leq 2$\ ,
~~~$r_1=0\ {\rm for}\ N>2$.

\nt{\bu{\bf bc}}~~~
$d ~=~  d^{bc} \ , \qquad r_{1}\leq 2$\ ,
~~~$r_{N-1}=0\  {\rm for}\ N>2$.

\nt{\bu{\bf bd}}~~~
$d ~=~  d^{bd} \ , \qquad r_1,r_{N-1}\leq 2$\ {\rm for}\ $N>2$,
~~~$1\leq r_1 \leq 4$\ {\rm for}\ $N=2$.

\newpage

\section{Reduction of supersymmetry in short and semi-short UIRs}
\label{sss}
\setcounter{equation}{0}

Our first task in this paper is to present explicitly the reduction of the
supersymmetries in the irreducible UIRs.  This means to give explicitly
the number ~$\k$~ of odd generators which are eliminated from the
 corresponding lowest weight module, (or equivalently, the number of super-derivatives that
 annihilate the corresponding  superfield).

\subsection{R-symmetry scalars}
\label{rsymscal}

We start with the simpler cases of $R$-symmetry scalars when  ~$r_i=0$~  for all $i$, which means also
that ~$m_1=m=m^*=0$.
These cases are valid also for $N=1$.  More explicitly:

\eqnn{rsyma}
\buu{\bf a} && d ~=~   d^a_{\vert_{m=0}} ~=~ 2 +2j_2 +z ~>~ 2 +2j_1-z\ , ~~~j_1 ~{\rm arbitrary},\nn \\
&& \k ~=~ N + (1-N)\d_{j_2,0}\ ,  ~~~{\rm or ~casewise:} \\
   && \k=N, ~~~{\rm if} ~~j_2>0 ,\nn\\   %~$\quarter$-BPS states,
   && \k=1,  ~~~{\rm if} ~~j_2=0 \nn\eea  %~$\quartern$-BPS states,
Here, ~$\k$~ is the number of anti-chiral generators ~$X^+_{3,5+N-k}$, $k=1,\ldots,\k$, ~ that are
eliminated. Thus, in the cases when ~$\k=N$~
the semi-short UIRs may be called semi-chiral
since they lack half of the anti-chiral generators.

\eqnn{rsymb} \buu{\bf b}
&& d ~=~ d^b_{\vert_{m=0}}  ~=~ z ~>~ 2 +2j_1 -z\ , ~~~j_1 ~{\rm arbitrary}, ~~
 j_2=0, \nn \\
 && \k=2N \  \eea  %~$\half$-BPS states
These short UIRs may be called chiral
since they lack all  anti-chiral generators ~$X^+_{3,4+k}\,$, ~$X^+_{4,4+k}\,$,
~$k=1,\ldots,N$.

\eqnn{rsymc}
\buu{\bf c} && d ~=~   d^c_{\vert_{m=0}} ~=~ 2 +2j_1 -z ~>~ 2 +2j_2 +z\ , ~~~j_2 ~{\rm arbitrary},\nn \\
&& \k ~=~ N + (1-N)\d_{j_1,0}\ ,  ~~~{\rm or ~casewise:} \\
   && \k=N, ~~~j_1>0 ,\nn\\   %~$\quarter$-BPS states,
   && \k=1,  ~~~j_1=0 \nn\eea %~$\quartern$-BPS states,
Here, ~$\k$~ is the number of chiral generators ~$X^+_{1,4+k}$, $k=1,\ldots,\k$, ~ that are
eliminated. Thus, in the cases when ~$\k=N$~
the semi-short UIRs may be called semi--anti-chiral
since they lack half of the chiral generators.

\eqnn{rsymd} \buu{\bf d}
&& d ~=~ d^d_{\vert_{m=0}}  ~=~ -z  ~>~ 2 +2j_2 +z \ , ~~~j_2 ~{\rm arbitrary}, ~~
 j_1=0, \nn \\
 && \k=2N    \eea  %~$\half$-BPS states
These short UIRs may be called anti-chiral
since they lack all  chiral generators ~$X^+_{1,4+k}\,$, ~$X^+_{2,4+k}\,$,
~$k=1,\ldots,N$.

\eqnn{rsymac}
\buu{\bf ac} &&  d ~=~ d^{ac}_{\vert_{m=0}}~=~  2 + j_1 + j_2 \ ,~~z ~=~ j_1-j_2\ ,\nn\\
&&\k ~=~ 2N + (1-N)(\d_{j_1,0}+\d_{j_2,0}), ~~~{\rm or ~casewise:} \\
&&\k=2N,  ~{\rm if} ~~ j_1,j_2>0 ,\nn\\  %~\half-BPS states,
&&\k=N+1,  ~{\rm if} ~~ j_1 >0,\ j_2=0 ,\nn\\
&&\k=N+1,  ~{\rm if} ~~ j_1=0,\ j_2>0 ,\nn\\
&&\k=2,  ~{\rm if} ~~ j_1=j_2=0 .\nn\eea
Here, ~$\k$~ is the number of mixed elimination: chiral generators ~$X^+_{1,4+k}$,
($k=1,\ldots,N + (1-N)\d_{j_1,0}$),
and anti-chiral generators ~$X^+_{3,5+N-k}$, ($k=1,\ldots,N + (1-N)\d_{j_2,0}$). Thus, in the cases when ~$\k=2N$~
the semi-short UIRs may be called semi--chiral--anti-chiral
since they lack half of the chiral and half of the anti-chiral generators.
(They may be  called Grassmann-analytic following \cite{FSa}.)

\eqnn{rsymad}
{\buu{\bf ad}} ~~~
&& d ~=~ d^{ad}_{\vert_{m=0}}   ~=~ 1 + j_2 ~=~ -z\ , ~~ ~j_1=0 ,  \nn\\
&&\k ~=~ 3N + (1-N)\d_{j_2,0}, ~~~{\rm or ~casewise:} \\
 &&\k=3N, ~~~ j_2>0 ,\nn\\  %~\tquarter-BPS states,
   && \k=2N+1,   ~~~ j_2=0.\nn\eea
Here, ~$\k$~ is the number of mixed elimination: both types chiral generators ~$X^+_{1,4+k}\,$, ~$X^+_{2,4+k}\,$,
($k=1,\ldots,N$),
and  anti-chiral generators ~$X^+_{3,5+N-k}$, ($k=1,\ldots,N + (1-N)\d_{j_2,0}$). Thus, in the cases when ~$\k=3N$~
the semi-short UIRs may be called  semi--chiral and anti-chiral
since they lack all the chiral and  half of the anti-chiral generators.

\eqnn{rsymbc}
{\buu{\bf bc}} &&
d ~=~ d^{bc}_{\vert_{m=0}}  ~=~ 1 + j_1 ~=~ z\ ,    ~~j_2=0,  \nn\\
&&\k ~=~ 3N + (1-N)\d_{j_1,0}, ~~~{\rm or ~casewise:} \\
 &&\k=3N,  ~~~ j_1>0 ,\nn\\  %~\tquarter-BPS states,
   &&\k=2N+1,   ~~~ j_1=0\ . \nn\eea
Here, ~$\k$~ is the number of mixed elimination:  chiral generators ~$X^+_{1,4+k}\,$, ($k=1,\ldots,N + (1-N)\d_{j_1,0}$)
 and both types anti-chiral generators ~$X^+_{3,4+k}$, ~$X^+_{4,4+k}\,$, ($k=1,\ldots,N$). Thus, in the cases when ~$\k=3N$~
the semi-short UIRs may be called  chiral and semi--anti-chiral
since they lack half of the chiral and  all of the anti-chiral generators.

 \md

The last two cases (ad,bc) form  two of the three series of massless states, holomorphic and antiholomorphic
\cite{DPu}, see also \cite{DPf,Doch}.

The case ~\bu{\bf bd}~ for $R$-symmetry scalars is trivial,
since also all other quantum numbers are zero ($d=j_1=j_2=z=0$).

\subsection{R-symmetry non-scalars}
\label{rsymns}

Here we need some additional notation. Let $N>1$ and let ~$i_0$~ be an integer such that
~$0\leq i_0\leq N-1\,$,  ~$r_i=0$~ for ~$i\leq i_0\,$,
and if ~$i_0<N-1$~ then ~$r_{i_0+1}> 0$.  Let now ~$i'_0$~ be an integer such that
~$0\leq i'_0\leq N-1\,$,  ~$r_{N-i}=0$~ for ~$i\leq i'_0\,$,
and if ~$i'_0<N-1$~ then ~$r_{N-1-i'_0}> 0$.\footnote{Both definitions are
formally valid for ~$N=1$~ with ~$i_0=0$~ since ~$r_{0}\equiv 0$~ by
convention and with ~$i'_0=0$~ since ~$r_{N}\equiv 0$~ by
convention.}

With this notation the cases of $R$-symmetry scalars occur when ~$i_0+i'_0=N-1$, thus,
from now on we have the restriction:
\eqn{} 0\leq i_0+i'_0\leq N-2  \ee

Now we can make a list  for the values of ~$\k$, with the same interpretation as in the previous
subsection, only the last case is added here.

\eqnn{rsymna}
\buu{\bf a} &&
d ~=~  d^a = 2 +2j_2 +z+2m_1 -2m/N > 2+2j_1 -z +2m/N \ ,\nn\\ &&  ~~j_1,j_2 ~{\rm arbitrary}, \nn\\
&& \k = 1 + i_0(1-\d_{j_2,0}) \leq N-1 \ .\eea
Here are eliminated the anti-chiral generators  ~$X^+_{3,5+N-k}\,$, $k\leq \k \,$.

\eqnn{rsymnb}
{\buu{\bf b}} &&
d ~=~ d^b = z   +2m_1 -2m/N > 2+2j_1 -z +2m/N \ ,   \nn\\ && ~~ j_2=0\ , ~~j_1 ~{\rm arbitrary}, \nn\\
&& \k = 2 +2i_0 \leq 2N-2 \ .\eea
Here are eliminated the anti-chiral generators  ~$X^+_{3,5+N-k}\,$, ~$X^+_{4,5+N-k}\,$,  $k\leq 1+i_0  \,$.

\eqnn{rsymnc}
\buu{\bf c} &&
d ~=~  d^c = 2+2j_1 -z +2m/N > 2 +2j_2 +z+2m_1 -2m/N \ , \nn\\ && ~~j_1,j_2 ~{\rm arbitrary}, \nn\\
&& \k = 1 + i'_0(1-\d_{j_1,0}) \leq N-1\ .\eea
Here are eliminated the chiral generators  ~$X^+_{1,4+k}\,$, $k\leq \k \,$.

\eqnn{rsymnd}
{\buu{\bf d}} &&
d ~=~ d^d = 2m/N  -z > 2 +2j_2 +z+2m_1 -2m/N  \ , \nn\\ && ~~ j_1=0, ~j_2 ~{\rm arbitrary},  \nn\\
&& \k = 2 +2i'_0 \leq 2N-2 \ .\eea
Here are eliminated the chiral generators  ~$X^+_{1,4+k}\,$, ~$X^+_{2,4+k}\,$,  $k\leq 1+i'_0  \,$.

\eqnn{rsymnac}
\buu{\bf ac}  &&
d ~=~  d^{ac} \ ,
 ~~~~z ~=~ j_1-j_2 +2m/N -m_1\ ,
~~j_1,j_2 ~{\rm arbitrary}, \nn\\
&& \k = 2 +i_0(1-\d_{j_2,0})+i'_0(1-\d_{j_1,0}) \leq N \ .\eea
Here  are eliminated  chiral generators  ~$X^+_{1,4+k}\,$, $k\leq 1+i'_0(1-\d_{j_1,0})\,$,
and anti-chiral generators  ~$X^+_{3,5+N-k}\,$, $k\leq 1+i_0(1-\d_{j_2,0})\,$.

\eqnn{rsymnad}
\buu{\bf ad}  &&
d ~=~  d^{ad} \ ,  ~~~ j_1=0\ ,
 ~~~z ~=~ 2m/N -m_1-1-j_2\,,~~   j_2 ~{\rm arbitrary},  \nn\\
&& \k = 3 +i_0(1-\d_{j_2,0})+2i'_0 \leq 1+N+i'_0 \leq 2N-1 \ .\eea
Here  are eliminated chiral generators  ~$X^+_{1,4+k}\,$, $X^+_{2,4+k}\,$, $k\leq 1+i'_0\,$,  and
anti-chiral generators  ~$X^+_{3,5+N-k}\,$, $k\leq 1+i_0(1-\d_{j_2,0})\,$.

\eqnn{rsymnbc}
\buu{\bf bc}  &&
d ~=~  d^{bc} \ ,  ~~~ j_2=0\ , ~~~
  z ~=~ 2m/N -m_1 +1+j_1\,, ~~j_1 ~{\rm arbitrary},  \nn\\
&&\k = 3 +2i_0+i'_0(1-\d_{j_1,0}) \leq 1+N+i_0 \leq 2N-1 \ .\eea
Here   are eliminated  chiral generators  ~$X^+_{1,4+k}\,$,  $k\leq 1+i'_0(1-\d_{j_1,0})\,$,
and anti-chiral generators  ~$X^+_{3,5+N-k}\,$,  ~$X^+_{4,5+N-k}\,$, $k\leq 1+i_0\,$.

\eqnn{rsymnbd}
\buu{\bf bd}  &&
d ~=~  d^{bd} ~=~ m_1 \ , ~ ~~ j_1=j_2=0\  , ~~~z = 2m/N -m_1\ , \nn\\
&& \k = 4 +2i_0+2i'_0 \leq   2N \ .\eea
Here  are eliminated  chiral generators  ~$X^+_{1,4+k}\,$, ~$X^+_{2,4+k}\,$, $k\leq 1+i'_0\,$,
and anti-chiral generators  ~$X^+_{3,5+N-k}\,$, ~$X^+_{4,5+N-k}\,$,  $k\leq 1+i_0\,$.\\
Note that the case ~$\k=2N$~ is possible exactly when ~$i_0+i'_0 = N-2$, i.e., when
 there is  only one nonzero ~$r_i$, namely, ~$r_{i_0+1}\neq 0$,  ~$i_0=0,1,\ldots,N-2$:
\eqn{rsymnbdd}
\buu{\bf bd} ~~~\k=2N  ~:~
d ~=~ m_1 ~=~ r_{i_0+1}\ , ~~ j_1=j_2=0\  , ~~z =   r_{i_0+1}\frac{2+2i_0-N}{N}\ . \ee
When ~$d=m_1=1$~ these $\ha$-eliminated UIRs form the 'mixed'
series of massless representations \cite{DPu}, see also \cite{DPf,Doch}.\footnote{This
series is absent for $N=1$.}\\

\nt{\bf Remark:}~~
In this paper we use the Verma (factor-)module realization of the
UIRs. We give here a short remark on what happens with the ER
realization of the
UIRs. As we know, cf. \cite{DPf}, the ERs are superfields depending on Minkowski
space-time and on $4N$ Grassmann coordinates ~$\theta_a^i$,
~$\bar{\theta}_b^k$, $a,b=1,2$, $i,k=1,\ldots,N$.
There is 1-to-1 correspondence in these dependencies
and the odd null conditions. Namely, if the condition
~$X^+_{a,4+k}\ |\L\rg ~=~ 0$, $a=1,2$, holds, then the superfields of the
corresponding ER do not depend on the variable ~$\theta_a^k\,$,
while if the condition
~$X^+_{a,5+N-k}\ |\L\rg ~=~ 0$, $a=3,4$, holds, then the superfields of the
corresponding ER do not depend on the variable
~$\bar{\theta}_{a-2}^k\,$.
These statements were used in the proof of unitarity for the ERs
picture, cf. \cite{DPp}, but were not explicated. They were analyzed in
detail in the papers \cite{AFSZ,FS,FSb,FSa}, using the notions of
'harmonic superspace analyticity' and Grassmann analyticity.\dia

{}In the next Section we shall  use the above classification  to the so-called BPS states.

\newpage

\section{BPS and possibly protected states}
\setcounter{equation}{0}

\subsection{PSU(2,2/4)}

The most interesting case is when ~$N=4$. This is related to super-Yang-Mills and contains the so-called BPS states,
cf., \cite{AFSZ,FSb,FSa,EdSo,Ryz,DHRy,ArSo,DHHR,Dolan}.
They are characterized by the number ~$\k$~ of odd generators which
annihilate them - then the corresponding state is called
~${\k\over 4N}\,$-BPS state.
Group-theoretically the case $N=4$  is special since the ~$u(1)$~
subalgebra carrying the quantum number $z$ becomes central
and one can invariantly set ~$z=0$.

We  give now the explicit list of these states:

\md

\nt\bu{\bf a} ~~~
$d ~=~   d^1_{41} ~=~ 2 +2j_2 +2m_1 -\ha m ~>~ d^3_{44}$ \ . ~~The last inequality  leads to the restriction:
 \eqn{resa} 2j_2 + r_1 > 2j_1 + r_3\ . \ee
In the case of $R$-symmetry scalars, i.e., $m_1=0$, follows that ~$j_2>j_1\,$, i.e., ~$j_2>0$, and then we have:
\eqn{bpsa} \k=4, ~~~~m_1=0, ~j_2>0 \ . \ee
In the case of $R$-symmetry non-scalars, i.e., $m_1\neq 0$, we have the range: ~$i_0+i'_0 \leq 2 $, and thus:
\eqn{bpsan} \k = 1 + i_0(1-\d_{j_2,0}) \leq  3\ . \ee

\nt{\bu{\bf b}}~~~
$d ~=~ d^2_{41} ~=~  \ha m^* ~>~ d^3_{44}\ , ~~ j_2=0$\ . ~~The last inequality  leads to the restriction:
\eqn{resb}  r_1 > 2 + 2j_1 + r_3 \ . \ee
The latter means that ~$r_1>2$, i.e., $m_1\neq 0$, ~$i_0=0$, and thus:
\eqn{bpsbn} \k = 2 \ . \ee

The next two cases are conjugate to the previous two so we present them shortly:

\nt\bu{\bf c} ~~~
$d ~=~   d^3_{44} ~=~ 2 +2j_1 + \ha m  ~>~ d^1_{41}\, ~~~\Ra$
\eqn{resc}  2j_1 + r_3 > 2j_2 + r_1  \ , \ee
$m_1=0 ~~\Ra ~~ j_1>j_2 ~~\Ra ~~ j_1>0 ~~\Ra$
\eqn{bpsc} \k=4, ~~ ~m_1=0, ~j_1>0\ . \ee
$m_1\neq 0 ~~\Ra~~ i_0+i'_0 \leq 2 ~~\Ra $
\eqn{bpscn} \k = 1 + i'_0(1-\d_{j_1,0}) \leq  3\ . \ee

\nt{\bu{\bf d}}~~~
$d ~=~ d^4_{44} ~=~ \ha m ~>~ d^1_{41}\,, ~~ j_1=0, ~~~\Ra$
\eqn{resd} r_3 > 2 + 2j_2 + r_1 \ ,\ee
$\Ra ~~ r_3 >2 ~~\Ra ~~ m_1\neq 0, ~~i'_0=0 ~~\Ra$
\eqn{bpsdn} \k ~=~ 2 \ . \ee

\nt\bu{\bf ac} ~~~
$d ~=~  d^{ac}  ~=~ 2 + j_1 + j_2 + m_1$\ . From $z=0$ follows:
\eqn{resac}  2j_2 + r_1 = 2j_1 + r_3 \ . \ee
In the case of $R$-symmetry scalars, i.e., $m_1=0$, follows that ~$j_2=j_1=j\,$,  and then we have:
\eqn{bpsac} \k ~=~ 8 -6 \d_{j,0} \ , \quad  d= 2+2j \ .\ee
In the case of $R$-symmetry non-scalars, i.e., $m_1\neq 0$,  ~$i_0+i'_0 \leq 2 $, and thus:
\eqn{bpsacn} \k = 2 +i_0(1-\d_{j_2,0})+i'_0(1-\d_{j_1,0}) \leq 4 \ . \ee

\nt{\bu{\bf ad}}~~~   From $z=0$ follows:  ~$  r_3 = 2 + 2j_2 +  r_1
~~\Ra ~~ r_3 \geq 2 ~~\Ra~~ m_1\neq 0$, and ~$i'_0=0 , ~~i_0\leq 2~~\Ra$
\eqnn{bpsadn}\k &=& 3 + i_0(1-\d_{j_2,0}) \leq 5  \ ,\nn\\
d ~&=&~ d^{ad} ~=~ 1 + j_2 + m_1 ~=~ 3 + 3j_2 + 2r_1 +r_2 \ ,  \\
\chi_4 &=& \{\,0\, ;\, r_1,r_2,2 + 2j_2 +  r_1\,;\,2j_2\,\}
\ .\nn \eea

\nt{\bu{\bf bc}}~~~   From $z=0$ follows:  ~$  r_1 = 2 + 2j_2 +  r_3
~~\Ra ~~ r_1 \geq 2 ~~\Ra~~ m_1\neq 0$, and ~$i_0=0 , ~~i'_0\leq 2~~\Ra$
\eqnn{bpsbcn}\k &=& 3 + i'_0(1-\d_{j_1,0}) \leq 5  \ ,\nn\\
d ~&=&~ d^{bc} ~=~ 1 + j_2 + m_1 ~=~ 3 + 3j_2 + 2r_1 +r_2 \ ,  \\
\chi_4 &=& \{\,2j_1\, ;\, 2 + 2j_2 +  r_3,r_2,r_3\,;\,0\,\}
\ .\nn \eea

\nt{\bu{\bf bd}}~~~From ~$z=0$~ follows:
 ~$ m=m^* ~\Rightarrow~ r_1 =  r_3 = r$,  thus, ~$i_0=i'_0 = 0,1$~ and then we have:
 \eqnn{bpsbdn} &&
 \k = 4 (1+i_0) \ , \\ && d ~=~ d^{bd} ~=~ m_1 ~=~ 2r +r_2 ~\neq~ 0 \ , ~~~r,r_2\in\bbz_+\ ,\nn\\
 &&\chi_4 = \{\,0\, ;\, r,r_2,r\,;\,0\,\} \ . \nn \eea

Some of these BPS-cases are extensively studied in the literature,
mostly those listed here as cases {\bf ac,bd}, cf.
\cite{AFSZ,FSb,FSa,EdSo,Ryz,DHRy,ArSo,DHHR,Dolan}.

\md

{}From the above BPS states we list now the most interesting ones in
 Tables 1-3:

\vspace{10mm}

\vbox{ \offinterlineskip
 %\centerline{
 {\bf Table 1}
\vskip0.3truecm
%\noindent
%\centerline{
$PSU(2,2/4)$, $\ \ha$-BPS states, ($\kappa=8$) \vskip0.3truecm
\halign{\baselineskip12pt \strut #& \vrule#\hskip0.1truecm & #\hfil&
\vrule#\hskip0.1truecm & #\hfil& \vrule#\hskip0.3truecm & #\hfil&
\vrule#\hskip0.1truecm & #\hfil& \vrule#\hskip0.1truecm & #\hfil&
\vrule#\hskip0.1truecm & #\hfil&
#\cr \tablerule %&&&& &&&& &&  \cr
&& case  &&\hskip 3mm $d$  && $j_1,j_2$ && $\ r_1,r_2,r_3$  &&
protected\ && \cr \tablerule &&&& &&&& && && \cr && {\bf ac}  &&  $\
2 + 2j\geq 3$  && $j=j_1=j_2\geq \ha$  && $\ m_1=0$  &&  &&     \cr
&&&& &&&& && && \cr \tablerule &&&& &&&& && && \cr && {\bf bd} && $\
r_2    \geq 1$&& $j_1=j_2= 0$&& $\ m_1 =  r_2$&&  &&\cr &&&& &&&& &&
&& \cr \tablerule
%&&&& &&&& &&  \cr
}}

\vspace{15mm}

\vbox{ \offinterlineskip
 \centerline{\bf Table 2}
\vskip0.3truecm
%\noindent
\centerline{ $PSU(2,2/4)$, $\ \quarter$-BPS states, ~($\k=4$)}
\vskip0.3truecm \halign{\baselineskip12pt \strut #&
\vrule#\hskip0.1truecm & #\hfil& \vrule#\hskip0.1truecm & #\hfil&
\vrule#\hskip0.3truecm & #\hfil& \vrule#\hskip0.1truecm & #\hfil&
\vrule#\hskip0.1truecm & #\hfil& \vrule#\hskip0.1truecm & #\hfil&
#\cr \tablerule %&&&& &&&& && && \cr
&&   &&\hskip 3mm $d$  && $j_1,j_2$ && $r_1,r_2,r_3$  && protected\
&&\cr \tablerule &&&& &&&& && && \cr && {\bf a} && $2 +2j_2 \geq 3$
&& $j_2\geq  \ha $&& $m_1= 0$&&  &&\cr &&&& &&&& && && \cr
\tablerule &&&& &&&& && && \cr && {\bf c} && $2 +2j_1 \geq 3$ &&
$j_1\geq  \ha $&& $m_1= 0$&&  &&\cr &&&& &&&& && && \cr
\tablerule
&&&& &&&& && && \cr
&& {\bf ac} && $2 + j_1+j_2 + r_{1+i_0}\geq 7/2$
&& $j_1-j_2 = \ha r_{1+i_0} (1-i_0)$, && $m_1=r_{1+i_0}> 0 ,$ &&
&&\cr
%&&&&  &&&& && && \cr
&&&& && $j_1+j_2\geq 1/2$ && $i_0 = 0,1,2$&&  &&\cr
%&&&&  &&&& && && \cr
\tablerule &&&& &&&& && && \cr && {\bf ad} && $\frac{m}{2}\geq
\frac{9}{2}$&& $j_1=0$, $j_2\geq \ha$ && $r_1=0$,&& No && \cr &&&&
&&  && $r_3=2 +2j_2$  && && \cr \tablerule &&&& &&&& && && \cr &&
{\bf bc} && $\frac{m^*}{2}\geq \frac{9}{2}$&& $j_1\geq \ha$, $j_2=0$
&& $r_1=2 +2j_1$,&& No && \cr &&&&  &&  && $r_3=0$  && && \cr
\tablerule &&&& &&&& && && \cr && {\bf bd} && $m_1  \geq
2$&&$j_1=j_2=0$&& $r_1=r_3\geq 1$&& No, if $r_1>2$ && \cr &&&& &&&&
&&  && \cr \tablerule
%&&&& &&&& && && \cr
}}

\vspace{15mm}

\vbox{ \offinterlineskip %\hrule
 \centerline{\bf Table 3}
\vskip0.3truecm
%\noindent
\centerline{ $PSU(2,2/4)$, $\ \eighth$-BPS states, ~($\k=2$)}
\vskip0.3truecm \halign{\baselineskip12pt \strut #&
\vrule#\hskip0.1truecm & #\hfil& \vrule#\hskip0.1truecm & #\hfil&
\vrule#\hskip0.3truecm & #\hfil& \vrule#\hskip0.1truecm & #\hfil&
\vrule#\hskip0.1truecm & #\hfil& \vrule#\hskip0.1truecm & #\hfil&
#\cr \tablerule %&&&& &&&& && && \cr
&& case  &&\hskip 3mm $d$  && $j_1,j_2$ && $r_1,r_2,r_3$  &&
protected\ &&\cr \tablerule &&&& &&&& && && \cr && {\bf a} && $2
+2j_2 + r_2 + \ha r_3$ &&$2j_2 > 2j_1 + r_3$ && $r_1=0, r_2 >0$ &&
&& \cr &&&&  &&&& && && \cr \tablerule &&&& &&&& && && \cr &&{\bf b}
&&  $\ha  m^* $ && $j_2=0$&& $r_1 > 2 + 2j_1 + r_3$ && No && \cr
&&&& &&&&  && && \cr \tablerule &&&& &&&& && && \cr && {\bf c} && $2
+2j_1 + r_2 + \ha r_1$ &&$2j_1 > 2j_2 + r_1$ && $r_3=0, r_2 >0$ &&
&& \cr &&&&  &&&& && && \cr \tablerule &&&& &&&& && && \cr &&{\bf d}
&&  $\ha  m $ && $j_1=0$&& $r_3 > 2 + 2j_2 + r_1$ && No && \cr &&&&
&&&&  && && \cr \tablerule &&&& &&&& && && \cr &&{\bf ac} && $2 +
m_1\geq 2$&& $j_1=j_2=0$ && && No, if $\ r_1r_3> 0 $ && \cr &&&&
&&&& &&  && \cr \tablerule
%&&&& &&&& && && \cr
}}

\vspace{10mm}

Finally, we remark that some of the above states would violate the
protectedness conditions that we gave in Subsection \ref{deco}. As
indicated in the last column of the above Tables these would be the
$\quarter\,$-BPS cases listed as cases {\bf ad,bc}, and in case {\bf
bd}   for $r_1=r_3>2$, while for the $\eighth\,$-BPS cases that
would be the cases {\bf b,d}, and in case {\bf ac} for $r_1r_3 >
0$.

%\newpage

\subsection{SU(2,2/N), ~$N=1,2$}

We can set $z=0$ also for $N\neq 4$ though this does not have the
same group-theoretical meaning as for $N=4$. In this Subsection we
treat separately the cases ~$N=1,2$, which are more peculiar.

\subsubsection{SU(2,2/1)}

For ~$N=1$~ setting $z=0$ is possible only for three cases ~{\bf a,c,ac}~  :

\nt{\bu{\bf a}} ~~$d= 2+ 2j_2$\ , ~~ $j_2>j_1\geq 0$,  \nlii
$\k=1$, ~~$\quarter\,$-BPS;

\nt{\bu{\bf c}} ~~$d= 2+ 2j_1$\ , ~~ $j_1>j_2\geq 0$,  \nlii
$\k=1$, ~~$\quarter\,$-BPS;

\nt{\bu{\bf ac}} ~~$d= 2+ 2j$\ , ~~ $j_1=j_2=j$,  \nlii
$\k=2$, ~~$\half\,$-BPS.

Note that according to the result of Subsection \ref{deco} the first two cases would not be protected.

\subsubsection{SU(2,2/2)}
\label{neq2}

For ~$N=2$~  holds ~$i_0=i'_0=0,1\,$.
Setting $z=0$ is possible for four cases ~{\bf a,c,ac,bd}~ when we have:

\nt{\bu{\bf a}} ~~$d= 2+ 2j_2 + r_1$\ , ~~ $j_2>j_1\geq 0$,  \nlii
$\k=1+i_0\leq 2$ ;

\nt{\bu{\bf c}} ~~$d= 2+ 2j_1 +r_1$\ , ~~ $j_1>j_2\geq 0$,  \nlii
$\k=1+i'_0\leq 2$ ;

\nt{\bu{\bf ac}} ~~$d= 2+ 2j +r_1$\ , ~~ $j_1=j_2=j$,  \nlii
$\k=4 -2\d_{i_0j,0}\leq 4$;

\nt{\bu{\bf bd}} ~~$d= r_1 >0$\ , ~~ $j_1=j_2=0$, (here $z=0$ holds
in all cases), \nlii $\k=4$, ~~$\half\,$-BPS.

Note that according to the result of Subsection \ref{deco} the first three cases would not be protected when $r_1>0 $,
i.e., when ~$i_0=i'_0=0$. In contradistinction, when $r_1= 0$,
i.e., ~$i_0=i'_0=1$, the first two are ~~$\quarter\,$-BPS, and the third, when $j>0$, a $\half\,$-BPS.
The fourth case would not be protected if ~$r_1>4$.

\subsection{SU(2,2/N), ~$N\geq 3$}

The  cases $N\geq 3$  are somewhat similar in these considerations
to $N=4$, (though some results differ), so we present them only in
Tables 4,5,6 and 7.

As we see the case of $\ \ha$-BPS states can be presented in a table
for all $N$.

The case of $\ \quarter$-BPS states for $N=3$ may be seen also in
the tables for general $N$, but  it makes sense to be presented
separately in Table 5.

\vspace{10mm}

\vbox{ \offinterlineskip %\hrule
 \centerline{\bf Table 4}
\vskip0.3truecm
%\noindent
\centerline{ $SU(2,2/N)$, $\ \ha$-BPS states, $\kappa=2N$, $N\geq 1$}
\vskip0.3truecm \halign{\baselineskip12pt \strut #&
\vrule#\hskip0.1truecm & #\hfil&
\vrule#\hskip0.1truecm & #\hfil&
\vrule#\hskip0.3truecm & #\hfil&
\vrule#\hskip0.1truecm & #\hfil&
\vrule#\hskip0.1truecm & #\hfil&
\vrule#\hskip0.1truecm & #\hfil&
#\cr \tablerule %&&&& &&&& &&  &&\cr
&&   &&\hskip 3mm $d$  && $j_1,j_2$ && $r_1,\ldots,r_{N-1}$  && protected\ && \cr
\tablerule
&&&& &&&& &&  &&\cr
&& {\bf ac}  &&  $2 + 2j\geq 3$  && $j=j_1= j_2\geq \ha$  && $m_1=0$  &&  &&     \cr
 &&  && &&&& &&  &&\cr
\tablerule
&&&& &&&& &&  &&\cr
&&{\bf bd} && $r_{\halfn} \geq 1$&& $j_1=j_2= 0$ && $m_1=r_{\halfn}$&& No, if $r_1>4$ &&\cr
&&$N$ even && &&&&    &&  for $N=2$&&  \cr
&&&& &&&& &&  &&\cr
\tablerule
%&&&& &&&& &&  \cr
}}

\vspace{10mm}

\vbox{ \offinterlineskip %\hrule
 \centerline{\bf Table 5}
\vskip0.3truecm
%\noindent
\centerline{ $SU(2,2/3)$, $\ \quarter$-BPS states, ~$\k=N=3$}
\vskip0.3truecm \halign{\baselineskip12pt \strut #&
\vrule#\hskip0.1truecm & #\hfil&
\vrule#\hskip0.1truecm & #\hfil&
\vrule#\hskip0.3truecm & #\hfil&
\vrule#\hskip0.1truecm & #\hfil&
\vrule#\hskip0.1truecm & #\hfil&
\vrule#\hskip0.1truecm & #\hfil&
#\cr \tablerule %&&&& &&&& &&&&  \cr
&&   &&\hskip 3mm $d$  && $j_1,j_2$ && $r_1,r_{2}$  && protected\ &&\cr
\tablerule
&&&& &&&& &&&&  \cr
&& {\bf a} && $2 +2j_2 \geq 3$ && $j_2\geq \ha $&& $m_1= 0$&&  &&\cr
&&&& &&&& &&&&  \cr
\tablerule
&&&& &&&& &&&&  \cr
&& {\bf c} && $2 +2j_1 \geq 3$ && $j_1\geq \ha $&& $m_1= 0$&&  &&\cr
&&&& &&&& &&&&  \cr
\tablerule
&&&& &&&& &&&&  \cr
&& {\bf ac} && $2 + j_1+j_2 + r_{1+i_0} \geq 6$ && $j_1-j_2 =\third (r_1-r_2) = $ && $m_1=r_{1+i_0}=$ &&  &&\cr
&&&& && $=\pm1,\pm2,\ldots$ && $=3,6,\ldots,$ &&  &&\cr
&&&& &&  && $i_0=0,1$ &&&&  \cr
&&&& &&&& &&&&  \cr
\tablerule
&&&& &&&& &&&&  \cr
&& {\bf ad} && $\frac{2}{3}m=\frac{2}{3}(r_1+2r_2)\geq 4$&&$j_1=0$, $j_2i_0=0$ && $r_2=3 +r_1+3j_2$ && No  && \cr
&&&& &&&& &&&&  \cr
\tablerule
&&&& &&&& &&&&  \cr
&& {\bf bc} && $\frac{2}{3}m^*=\frac{2}{3}(2r_1+r_2)\geq 4$&&$j_2=0$, $j_1i'_0=0$ && $r_1=3 +r_2+3j_1$ && No  && \cr
&&&& &&&& &&&&  \cr
\tablerule
}}

\vspace{10mm}
%\np %{5mm}

\vbox{ \offinterlineskip %\hrule
 \centerline{\bf Table 6}
\vskip0.3truecm
%\noindent
\centerline{ $SU(2,2/N)$, $\ \quarter$-BPS states, ~$\k=N$, $N>4$}
\vskip0.3truecm \halign{\baselineskip12pt \strut #&
\vrule#\hskip0.1truecm & #\hfil&
\vrule#\hskip0.1truecm & #\hfil&
\vrule#\hskip0.3truecm & #\hfil&
\vrule#\hskip0.1truecm & #\hfil&
\vrule#\hskip0.1truecm & #\hfil&
\vrule#\hskip0.1truecm & #\hfil&
#\cr \tablerule %&&&& &&&& &&&&  \cr
&&   &&\hskip 3mm $d$  && $j_1,j_2$ && $\ r_1,\ldots,r_{N-1}$  && protected\ &&\cr
\tablerule
 &&  &&   &&  &&   &&  &&   \cr
 && {\bf a}  &&  $2 +2j_2 $  &&  $j_2\geq \ha $ &&  $m_1= 0$ &&    && \cr
 &&  &&    &&  &&   &&  &&   \cr
 \tablerule
 &&  &&   &&  &&   &&  &&   \cr
 && {\bf c}  &&  $2 +2j_1 $  &&  $j_1\geq \ha  $ &&  $m_1= 0$ &&    && \cr
 &&  &&    &&  &&   &&  &&   \cr
\tablerule
 &&  &&   &&  &&   &&  &&   \cr
 &&  {\bf ac} &&  $2 + j_1+j_2 + r_{1+i_0}$  &&  $j_1-j_2 =$ &&  $m_1=r_{1+i_0} > 0,$  &&   && \cr
 &&   &&    && $r_{1+i_0} (1-\novert(1+i_0))$  &&  $i_0\leq N-2$  &&    && \cr
% &&  &&   &&  $$ &&   &&  &&   \cr
 &&  &&   &&  &&   &&  &&   \cr
\tablerule
 &&  &&   &&  &&   &&  &&   \cr
 && {\bf ad} &&  $1 + m_1$ && $j_1=j_2=0$  &&  $i'_0=\frac{N-3}{2}$ && No, if $r_1> 0 $ &&   \cr
 && $N$ odd &&   &&  &&   &&  &&   \cr
 &&  &&   &&  &&   &&  &&   \cr
\tablerule
 &&  &&   &&  &&   &&  &&   \cr
 && {\bf ad} &&  $\frac{2m}{N}$ && $j_1=0$, $j_2\geq \ha$  &&  $i_0+i'_0\leq N-3 $  &&  No, if $r_1> 0$ &&  \cr
 &&  && $  $  &&    &&  $ $   &&  or $r_{N-1}> 2$ &&  \cr
 &&  &&   &&  &&   &&   &&   \cr
 \tablerule
 &&  &&   &&  &&   &&  &&   \cr
 && {\bf bc} &&  $1 + m_1$ && $j_1=j_2=0$  &&  $i_0=\frac{N-3}{2}$ && No, if $r_{N-1}> 0 $ &&   \cr
 && $N$ odd &&   &&  &&   &&  &&   \cr
 &&  &&   &&  &&   &&  &&   \cr
\tablerule
 &&  &&   &&  &&   &&  &&   \cr
 && {\bf bc} &&  $\frac{2m^*}{N}$ && $j_1\geq \ha$, $j_2=0$  &&  $i_0+i'_0\leq N-3 $  &&  No, if $r_{N-1}> 0$ &&  \cr
 &&  && $  $  &&    &&  $ $   && or $r_1> 2$  &&  \cr
 &&  &&   &&  &&   &&  &&   \cr
\tablerule
 &&  &&   &&  &&   &&  &&   \cr
 && {\bf bd} &&  $m_1 >0 $  &&  $j_1=j_2=0$ &&  $i_0+i'_0=\frac{N}{2}-2$ &&  No, if  &&  \cr
 &&  $N$ even &&   &&  &&  $ $ &&  $r_1,r_{N-1} >2$  &&  \cr
 &&  &&   &&  &&   && $$ &&  \cr
\tablerule
% &&  &&   &&  &&   &&   &&  \cr
}}

\vspace{10mm}

\vbox{ \offinterlineskip %\hrule
 \centerline{\bf Table 7}
\vskip0.3truecm
%\noindent
\centerline{ $SU(2,2/N)$, $\ \eighth$-BPS states, ~$\k=N/2$, $N$ even, $N>4$}
\vskip0.3truecm \halign{\baselineskip12pt \strut #&
\vrule#\hskip0.1truecm & #\hfil&
\vrule#\hskip0.1truecm & #\hfil&
\vrule#\hskip0.3truecm & #\hfil&
\vrule#\hskip0.1truecm & #\hfil&
\vrule#\hskip0.1truecm & #\hfil&
\vrule#\hskip0.1truecm & #\hfil&
#\cr \tablerule %&&&& &&&& &&&&  \cr
&& case &&\hskip 3mm $d$  && $j_1,j_2$ && $\ r_1,\ldots,r_{N-1}$  && protected\ &&\cr
\tablerule
 &&  &&   &&  &&   &&  &&   \cr
 && {\bf a}  &&  $2 +2j_2 + \novert m^*$  &&  $j_1 -j_2  < \frac{m^*-m}{N}  $ && $i_0=\halfn-1$ &&    && \cr
 &&  &&    &&  &&   &&  &&   \cr
 \tablerule
 &&&& &&&& &&&&  \cr
  && {\bf b} &&   $\novert m^*$ && $j_1 +1  < \frac{m^*-m}{N}\,,  $    && $i_0= \quartern -1$&&  &&   \cr
&& $N\in 4\bbn$ &&    && $j_2=0$ &&   &&  &&   \cr
\tablerule
 &&  &&   &&  &&   &&  &&   \cr
 && {\bf c}  &&  $2 +2j_1 + \novert m$  &&  $ j_2 -j_1 < \frac{m-m^*}{N} $ && $i'_0=\halfn-1$ &&    && \cr
 &&  &&    &&  &&   &&  &&   \cr
 \tablerule
 &&&& &&&& &&&&  \cr
  && {\bf d} &&   $\novert m$ && $ j_2 + 1 < \frac{m-m^*}{N}\,, $    && $i'_0= \quartern -1$&&  &&   \cr
&& $N\in 4\bbn$ &&    && $j_1=0$ &&   &&  &&   \cr
\tablerule
&&  &&   &&  &&   &&  &&   \cr
 && {\bf ac}  &&  $2 +j_1+j_2 + m_1$  && $j_1 -j_2  = \frac{m^*-m}{N}  $, && $
i_0(1-\d_{j_2,0})+i'_0(1-\d_{j_1,0})$ &&    && \cr
 &&  &&    && $j_1+j_2>0$  && $= \halfn-2$  &&  &&   \cr
 \tablerule
 &&&& &&&& &&&&  \cr
  && {\bf ad} &&   $\novert m$ && $j_2 +1  = \frac{m-m^*}{N},$  && $i_0+2i'_0= \halfn -3$ && No, if $r_1>0$  &&   \cr
&& $N\geq 6$ &&    && $j_1=0$, $j_2>0$ &&   && or $r_{N-1}>2$ &&   \cr
\tablerule
&&&& &&&& &&&&  \cr
  && {\bf ad} &&   $\novert m$ && $j_1=j_2=0$ && $\frac{m-m^*}{N} = 1,$ $ i'_0= \ha(\halfn -3)$ && No, if $r_1>0$ &&   \cr
&& $N= 6,10,\ldots$ &&    &&  &&   &&  or $r_{N-1}>2$ &&   \cr
 \tablerule
 &&&& &&&& &&&&  \cr
  && {\bf bc} &&   $\novert m^*$ && $j_1 +1=\frac{m^*-m}{N} $  && $2i_0+i'_0= \halfn -3$ && No, if $r_1>2$ &&   \cr
&& $N\geq 6$ &&    && $j_1>0$, $j_2=0$ &&   &&  or $r_{N-1}>0$ &&   \cr
\tablerule
&&&& &&&& &&&&  \cr
  && {\bf bc} &&   $\novert m^*$ && $j_1=j_2=0$  && $\frac{m^*-m}{N} = 1,$ $ i_0= \ha(\halfn -3)$ && No, if $r_1>2$ &&   \cr
&& $N= 6,10,\ldots$ &&    &&  &&   && or $r_{N-1}>0$ &&   \cr
\tablerule
&&&& &&&& &&&&  \cr
  && {\bf bd} &&   $m_1$ &&  $j_1=j_2=0$ && $ m=m^*$, $ i_0+i'_0= \quartn -2$ && No, if &&   \cr
&& $N= 8,12,\ldots$ &&    &&  &&   && $r_1, r_{N-1}>2$ &&   \cr
 \tablerule}}

\vspace{15mm}
%\np

\section{Outlook}

In the present paper, we gave explicitly the  reduction of supersymmetries of the positive
energy unitary irreducible representations of the N-extended D=4
conformal superalgebras  su(2,2/N). Further we give the classification of BPS and
   possibly protected states.
Our considerations are group-theoretic and model-independent.
Thus, we could give only the necessary conditions for protectedness, or equivalently,  the
sufficient conditions for unprotectedness.

\section*{Acknowledgments}

The author would like to thank for hospitality the International
School for Advanced Studies, Trieste, and the Erwin Schr\"odinger
Institute, Vienna, where part of the work was done.  The author was
supported in part by Bulgarian NSF grant {\it DO 02-257}.

\end {document}